\title{S-hull: a fast radial sweep-hull routine for Delaunay triangulation}
\author{
        D. A. Sinclair \\
         {\tt david@s-hull.org}\\
         Cambridge, UK
}
\date{\today}

\documentclass[12pt]{article}
\usepackage{graphicx}
\begin{document}
\maketitle

\begin{abstract}
A new O(nlog(n)) algorithm is presented for performing Delaunay triangulation of sets of 2D points.
The novel component of the algorithm is a radially propagating \emph{sweep-hull} (sequentially created from 
the radially sorted set of 2D points, giving a non-overlapping triangulation),  
paired with a final triangle flipping step to give the Delaunay triangluation.

In empirical tests the algorithm runs in approximately half the time of q-hull for 2D Delaunay 
triangulation on randomly generated point sets.

\end{abstract}

\section{Introduction}
Delaunay triangulation \cite{Delaunay} is a generically useful technique for quickly generating 
non-overlapping triangular meshes that reflect the nearest neighbour structure of a set of points in 2D.
There now exist a number of O(n \emph{log}(n)) (or better) techniques for finding the Delaunay triangulation of a set of 2D points.
Such methods include q-hull \cite{Barber96thequickhull} which uses the property that the convex hull (in R(d+1) ) corresponds to the Delaunay triangulation in R(d), randomised point insertion algorithms as in \cite{insertion} and the divinely inspired sweepline algorithm \cite{sweepline}.

In this paper a hybred algorithm is presented that uses a radially propagating \emph{sweep-hull} over a radially 
sorted set of points in conjunction with a final triangle flipping step. 
As points are sequentially added and deleted from the sweep-hull, triangles are created and stored in a graph, 
to create a non-overlapping triangulation of the 2D point set.
In contrast to Fortune's algorithm only the current convex hull of the swept portion of the set needs to be maintained and searched for point insertion (rather than the more expensive event que).

\paragraph{Outline}
The following section provide a description of the S-hull algorithm and a performance comparison with q-hull.
GPL source for S-hull is available from {\tt http://s-hull.org}.

\section{The S-hull algorithm}\label{algorithm}

S-hull operates as follows:
For a set of unique points \( {\bf x}_i\) in R2:

\begin{enumerate}
\item sellect a seed point \({\bf x}_0\) from \({\bf x}_i\).
\item sort according to\( |{\bf x}_i - {\bf x}_0|^2\).
\item find the point \({\bf x}_j\) closest to \( {\bf x}_0\).
\item find the point \({\bf x}_k\) that creates the smallest circum-circle with \({\bf x}_0\) and \({\bf x}_j\)
  and record the center of the circum-circle {\bf C}.
\item order points \( [ {\bf x}_0, {\bf x}_j, {\bf x}_k ] \) to give a right handed system:
   this is the initial seed convex hull.
\item resort the remaining points according to \(|{\bf x}_i - {\bf C}|^2\) to give points \({\bf s}_i\).
\item sequentially add the points \({\bf s}_i\) to the porpagating 2D convex hull that is seeded with the triangle formed from \( [ {\bf x}_0, {\bf x}_j, {\bf x}_k ] \).
as a new point is added the facets of the 2D-hull that are {\emph visible} to it form new triangles.
\item a non-overlapping triangulation of the set of points is created. (This is an extremely fast method for creating an non-overlapping triangualtion of a 2D point set).
\item adjacent pairs of triangles of this triangulation must be 'flipped' to create a Delaunay triangulation
from the initial non-overlapping triangulation. 
\end{enumerate}

The algorithm generates a Delaunay triangulation together with the 2D convex hull for set of points.

Figure 1 shows a randomly generated set of 100 points in R2 with the initial triangular seed hull marked in red.
Figure 2 shows the propagation of the sweep-hull, new triangles in red, existing triangles in blue. 
Figure 3 shows the delaunay triangulation generated by s-hull.
Table 1 gives empirical times for s-hull and q-hull for point sets that range in size form 100 to 1,000,000.
S-hull is empirically faster. The test code was compiled and run on a MacBook Pro using gcc \tt{i686-apple-darwin9-g++-4.0.1}.

\begin{figure}[!htb] 
\includegraphics[width=\textwidth]{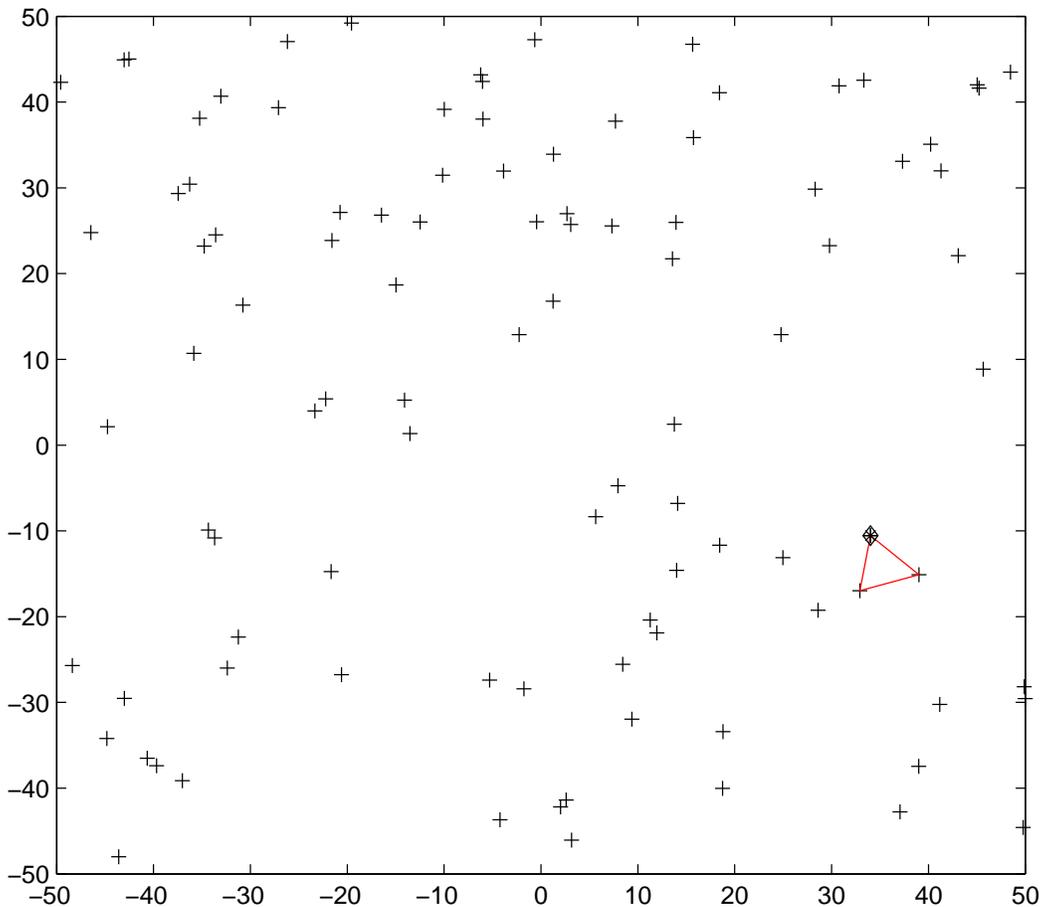}
\caption{ \emph{ Randomly generatesd set of points in R2. The seed point for the triangulation is marked along with the triangle associated with the smallest circum-circle through it and its nearest neighbour.}}
\label{Figure:Graphs1}
\end{figure}

\begin{figure}[!htb] 
\includegraphics[width=\textwidth]{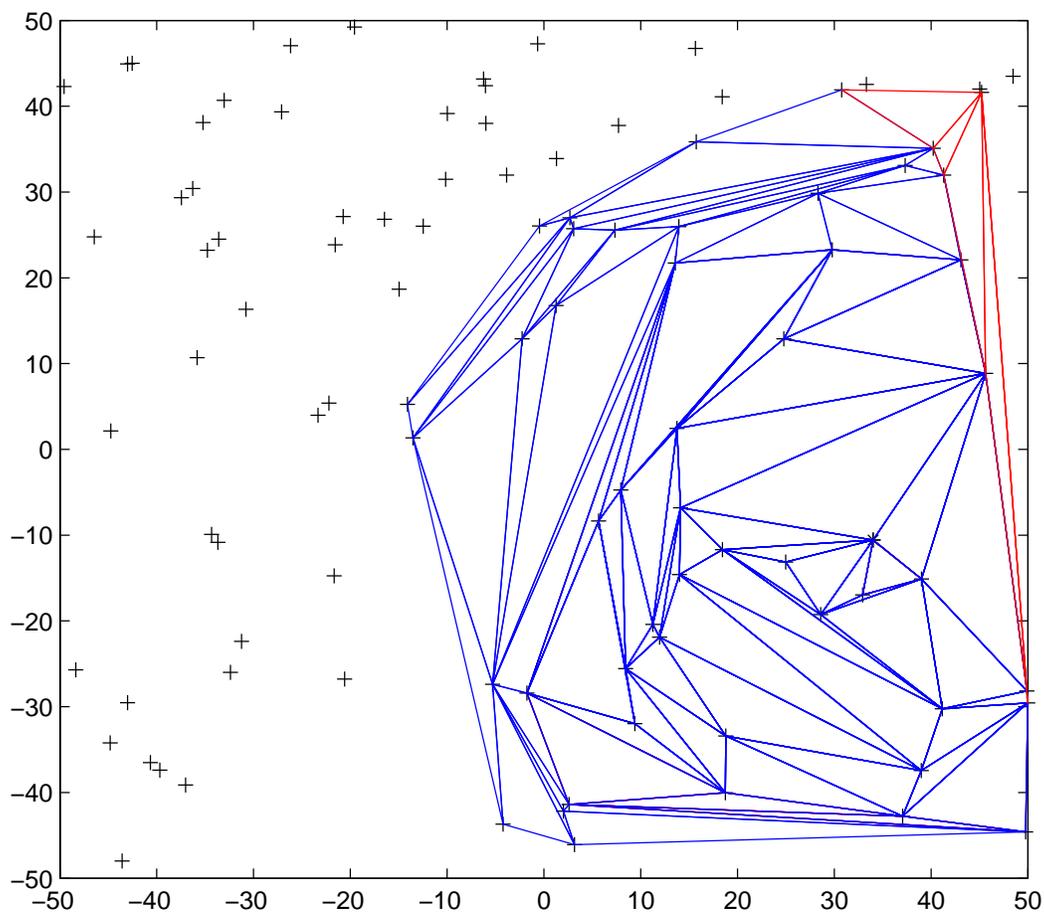}
\caption{ \emph{Randomly generatesd set of points in R2. Sequential addition of triangles as the convex hull is swept through the ordered set of points (new triangles in red).}}
\label{Figure:Graphs2}
\end{figure}

\begin{figure}[!htb] 
\includegraphics[width=\textwidth]{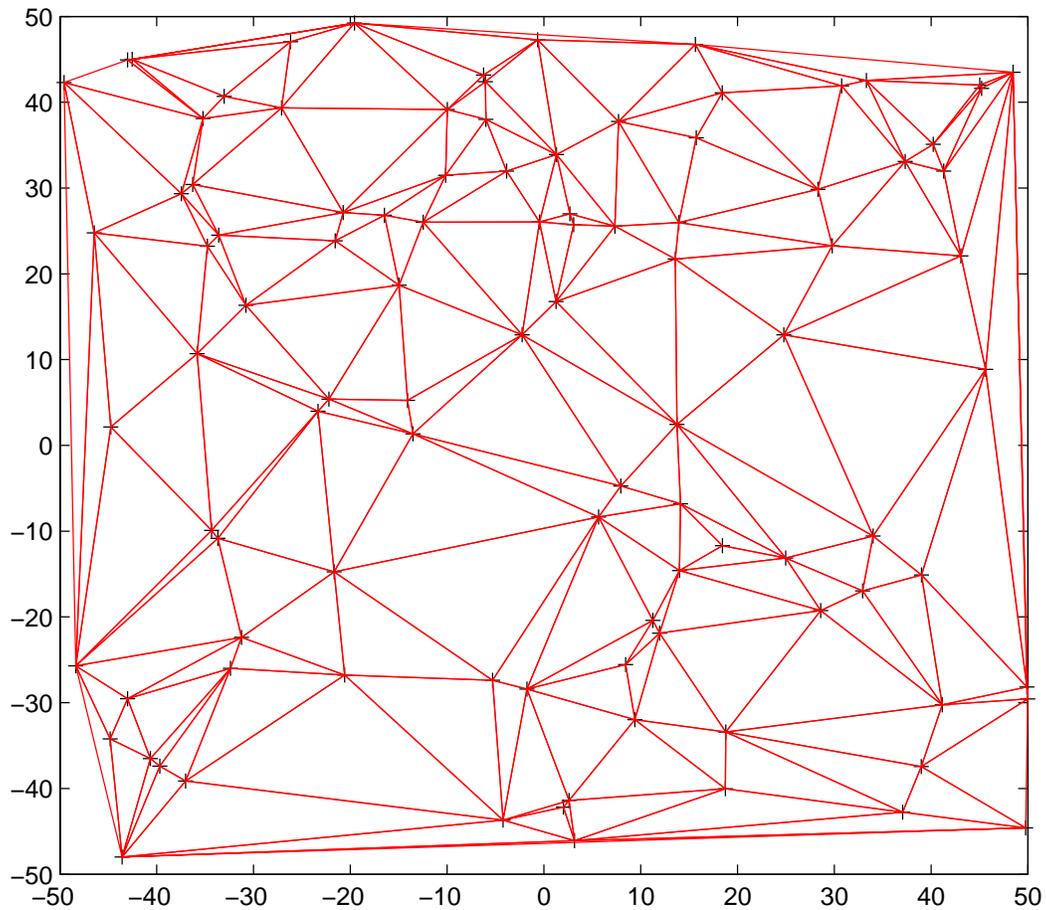}
\caption{\emph{Delaunay triangulation of the set of points in R2 generated by s-hull.}}
\label{Figure:Graphs3}
\end{figure}

\begin{table}[!ht]
\begin{center}
\begin{tabular}{|l|l|l|l|l|l|}
\hline
algorithm & 100 pts & 1000 & 10,000 & 100,000 & 1,000,000 \\
\hline
q-hull    & 0.001263 & 0.01154  & 0.1095 & 1.961 & 24.3 \\
s-hull    & 0.000751 & 0.005682 & 0.0766 & 1.044 & 14.4905\\
\hline
\end{tabular}
\end{center}
\caption{\emph{Timings in seconds for Delaunay triangualtion of 2D point sets in size range 100 - 1,000,000.}}
\label{Table:timings}
\end{table}

The improvement in speed over q-hull probably stems from the smaller size and lower cost of 
visible facet search of the 2D hull in S-hull verses the 3D hull used in q-hull (to directly generate 
the Delaunay triangulation).

\section{Implementation details}\label{coding}
A GPL version of S-hull in C++ is provided at  \tt{http://www.s-hull.org}.
The reference code provided uses the standard sort routine from the STL to perform the radial sorting of 
a set of points. This naturally returns the nearest neighbour of the seed point. In searching for the
point that is associated with the smallest circum circle containitng points \( \bf{x}_0 \) and \( \bf{x}_j \) points are searched 
in assending order of distance from \( \bf{x}_0 \) 
only up to the point where the current minimum diameter circum 
circle is smaller than the distance to the point.
The triangle graph generated retains the center of each triangle's circum-circle and the square of its radius
and this is used directly in the triangle flipping routine. 
The triangle flipping routine is applied iteratively until no triangles flip.

Roundoff error is an issue in determining whether a pair of triangles should be 'flipped'. In the case that the difference in minimum circum circle diameter for one confiiguration of a pair triangles verses the other is relatively small a double precision routine is used. A limit is placed on the number of times a pair of triangles may be flipped as a guard against degenerate cases.

\section{Conclusions}\label{conclude}
Empirically S-hull performs Delaunay triangluation substantially faster than q-hull.
It is however slower than q-hull for purely finding the convex hull of a set of 2D points and 
does not perform operations in higher dimensions.
The author believes that additional performace improvement would be possible through more efficient code.

\section{Acknowledgements}\label{acknow}
Steven Fortune's \emph{ sweep-line} algorithm is acknowledged as the inspiration for the \emph{sweep-hull} 
algorithm given here.
The author would like to acknowledge material support form RedGate.com a genuinely enlightened company.

\bibliographystyle{plain}
\bibliography{bib_lib}

\begin{thebibliography}{1}

\bibitem{Barber96thequickhull}
C.~Bradford Barber, David~P. Dobkin, and Hannu Huhdanpaa.
\newblock The quickhull algorithm for convex hulls.
\newblock {\em ACM TRANSACTIONS ON MATHEMATICAL SOFTWARE}, 22(4):469--483,
  1996.

\bibitem{Delaunay}
Boris~N. Delaunay.
\newblock Sur la sphere vide.
\newblock {\em Bulletin of Academy of Sciences of the USSR}, (6):793--800,
  1934.

\bibitem{sweepline}
S.~Fortune.
\newblock Stable maintenance of point-set triangulation in two-dimensions.
\newblock {\em 30th Annual Symposium on the Foundations of Computer Science.
  IEEE, New York.}, 1989.

\bibitem{insertion}
L.~Guibas, D.~Knuth, and M.~Sharir.
\newblock Randomized incremental construction of delaunay and voronoi diagrams.
\newblock {\em Algorithmica}, 7:381–413, 1992.

\end{thebibliography}

\end{document}